\documentclass{webofc}

\usepackage[varg]{txfonts}   
\usepackage{hyperref}
\usepackage{url}
\hypersetup{colorlinks=true,citecolor=blue,urlcolor=blue,linkcolor=blue}
\usepackage[rightcaption]{sidecap}
\usepackage{graphicx}

\newcommand{\pp}{\mbox{\textit{p+p}}\xspace}

\begin{document}

\title{\raggedright Energy dependence of $\phi$(1020) meson production in nucleus-nucleus collisions at the CERN SPS}
\author{
  \firstname{Łukasz}
  \lastname{Rozpłochowski}
  \inst{1}\fnsep\thanks{\email{lukasz.rozplochowski@ifj.edu.pl}}
  for NA61/SHINE Collaboration
}
\institute{
  H. Niewodniczański Institute of Nuclear Physics, PAS, Radzikowskiego 152, 31-342 Kraków, Poland
}

\abstract{
This paper presents preliminary results from the NA61/SHINE experiment on $\phi$
meson production in central Ar+Sc collisions at beam momenta of 150$A$, 75$A$,
and 40$A$ GeV/$c$ (corresponding to $\sqrt{s_{NN}}$ = 16.8, 11.9, and 8.8 GeV,
respectively). These results include double differential distributions in
rapidity ($y$) and transverse momentum ($p_T$), as well as $p_T$-integrated
rapidity distributions and total $\phi$ yields. Additionally, the widths of the
rapidity distributions of $\phi$ and $\phi/\pi$ yield ratios are compared to
those from Pb+Pb and \pp reactions. Notably, this work represents the first-ever
measurements of $\phi$ production in a mid-sized system at SPS energies.
}

\maketitle

\section{Introduction}
\label{intro}
The $\phi$ meson, a resonance particle with a rest mass $m_0 = 1019.461 \pm 0.016$ MeV and a mass width $\Gamma = 4.249 \pm 0.013$ MeV, consists almost entirely of strange and anti-strange quarks. As the lightest hidden-strangeness particle, it is valuable for studying the Quark-Gluon Plasma. In a hot and dense QGP environment, $\phi$ production may be particularly sensitive to strangeness enhancement effects, potentially behaving like a "doubly strange" particle.

The methods applied to investigate $\phi$ production in Ar+Sc reactions are largely consistent with those used in the analysis of $\phi$ production in \pp collisions~\cite{NA61SHINE:2019gqe, Marcinek:2234262}. Direct observation of the $\phi$ meson is not feasible experimentally, so the $K^+K^-$ decay channel is used as an indirect method to study $\phi$ production. The entire available phase space was divided into bins, and the tag-and-probe method~\cite{28a6c7dfce674dcb9b80b711fa851293} was used to extract the number of $\phi$ mesons in each bin.

\section{Results}
\label{results}
\begin{figure}[]
  \centering
  \includegraphics[page=2, width=1\textwidth]{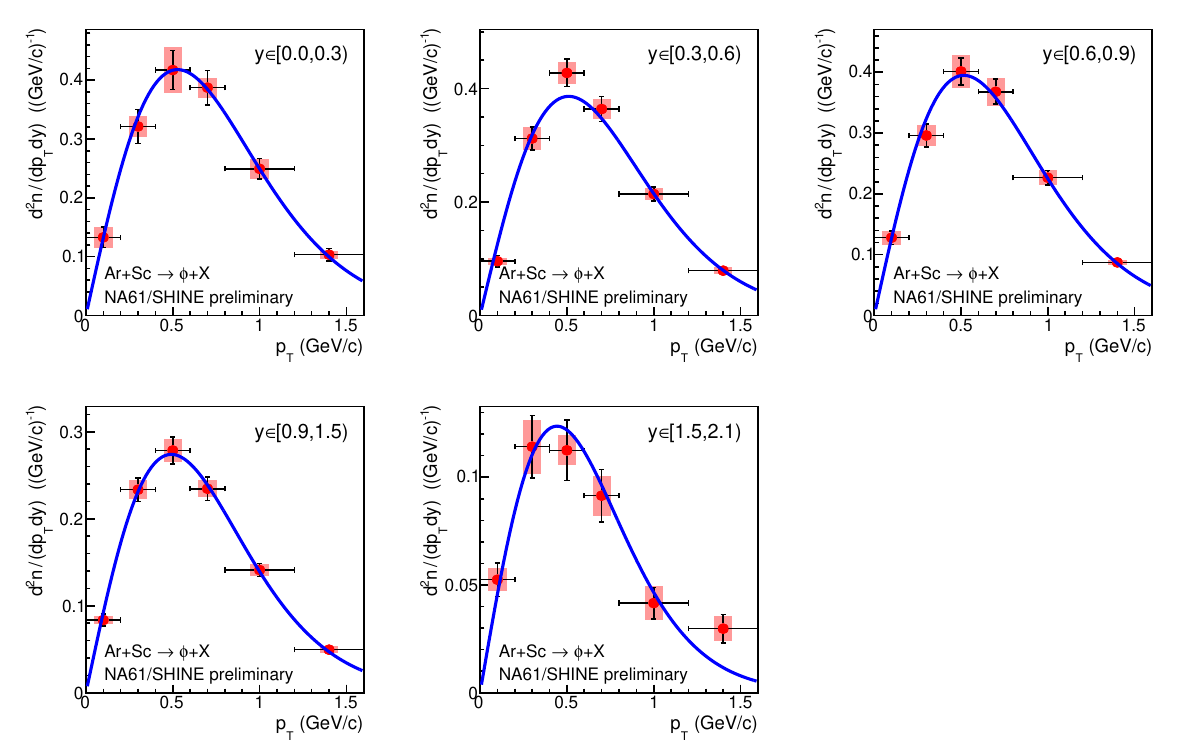}
  \caption{Double differential spectra ($y$, $p_T$) of $\phi$ produced in central Ar+Sc collisions at $p_{\text{beam}}$ = 150$A$~GeV/$c$. Statistical and systematic uncertainties are represented by vertical bars and red rectangles, respectively.  The blue curve is a fit used to obtain the integral of the unmeasured tail of the distribution.}
  \label{fig:d2dxFdpT}
\end{figure}

\begin{figure}[]
  \centering
  \includegraphics[page=2, width=0.32\textwidth]{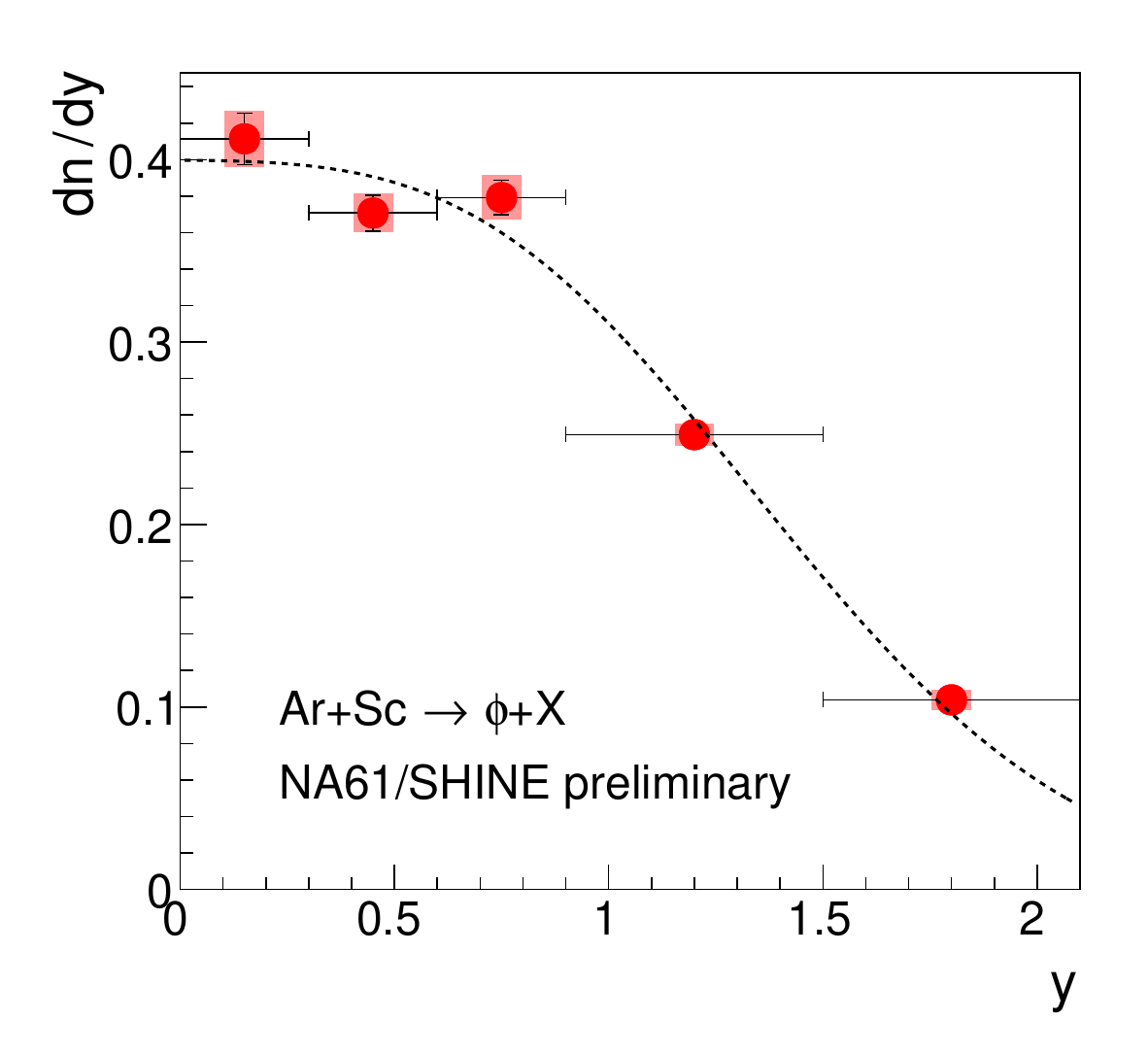}
  \includegraphics[page=2, width=0.32\textwidth]{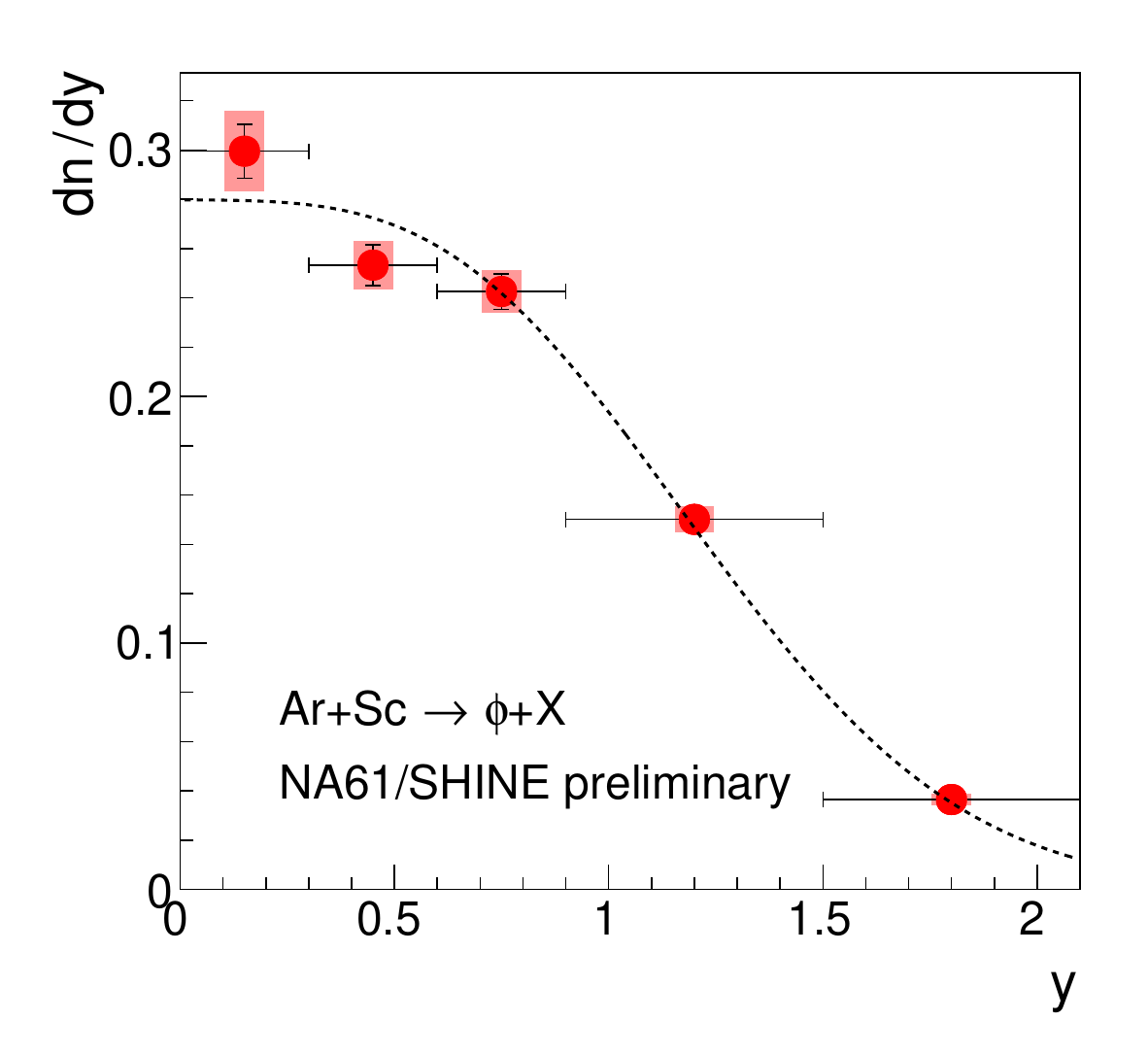}
  \includegraphics[page=2, width=0.32\textwidth]{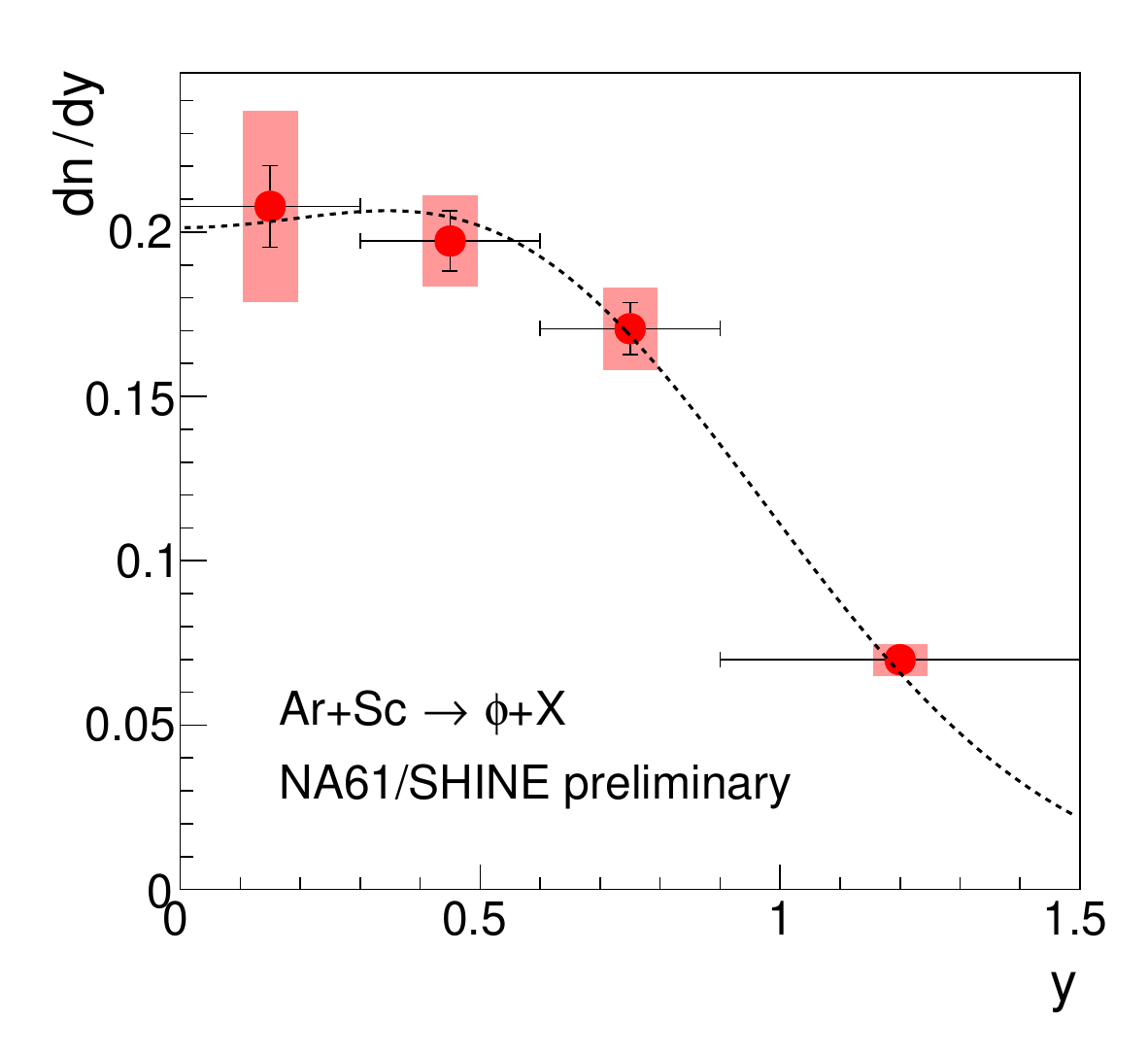}
  \caption{Rapidity distributions of $\phi$ mesons produced in central Ar+Sc collision
  at beam momenta equal to 150$A$, 75$A$ and 40$A$ GeV/$c$ (from left to right).
The dotted lines are the double Gaussian functions (see text) fitted to the experimental points. Statistical uncertainties are represented by vertical bars and systematic uncertainties by red rectangles.}
  \label{fig:rapidity}
\end{figure}

Fig.~\ref{fig:d2dxFdpT} shows double differential ($y$, $p_T$) distributions of $\phi$ mesons produced in central Ar+Sc collisions at 150$A$ GeV/$c$ beam momentum. Integrating these distributions over $p_T$, yields the rapidity distribution (Fig.~\ref{fig:rapidity}, left). A correction (up to 5\%) for the tail of the double differential distribution is calculated from the fit of the function $f(p_T) \propto p_T \cdot \exp(-m_T/T)$ visible as blue curves in Fig.~\ref{fig:d2dxFdpT}.

Rapidity distributions for all three energies considered here, are presented in
Fig.~\ref{fig:rapidity}. These distributions are fitted with the sum of two Gaussian functions:
\begin{equation}
  g(\textrm{y}) \propto \exp \Big( \frac{-(x-\mu)^2}{2\sigma^2} \Big) + \exp \Big( \frac{-(x+\mu)^2}{2\sigma^2} \Big)\,.
\end{equation}
From these fits, the widths RMS = $\sqrt{\sigma^2 + \mu^2}$ of the distributions are obtained, and the corrections (up to 2.5\%) for the tails of the spectra. By integrating the rapidity distributions, one determines the total yields $\langle
\phi \rangle$.

\begin{SCfigure}
  \includegraphics[page=4,width=0.5\textwidth]{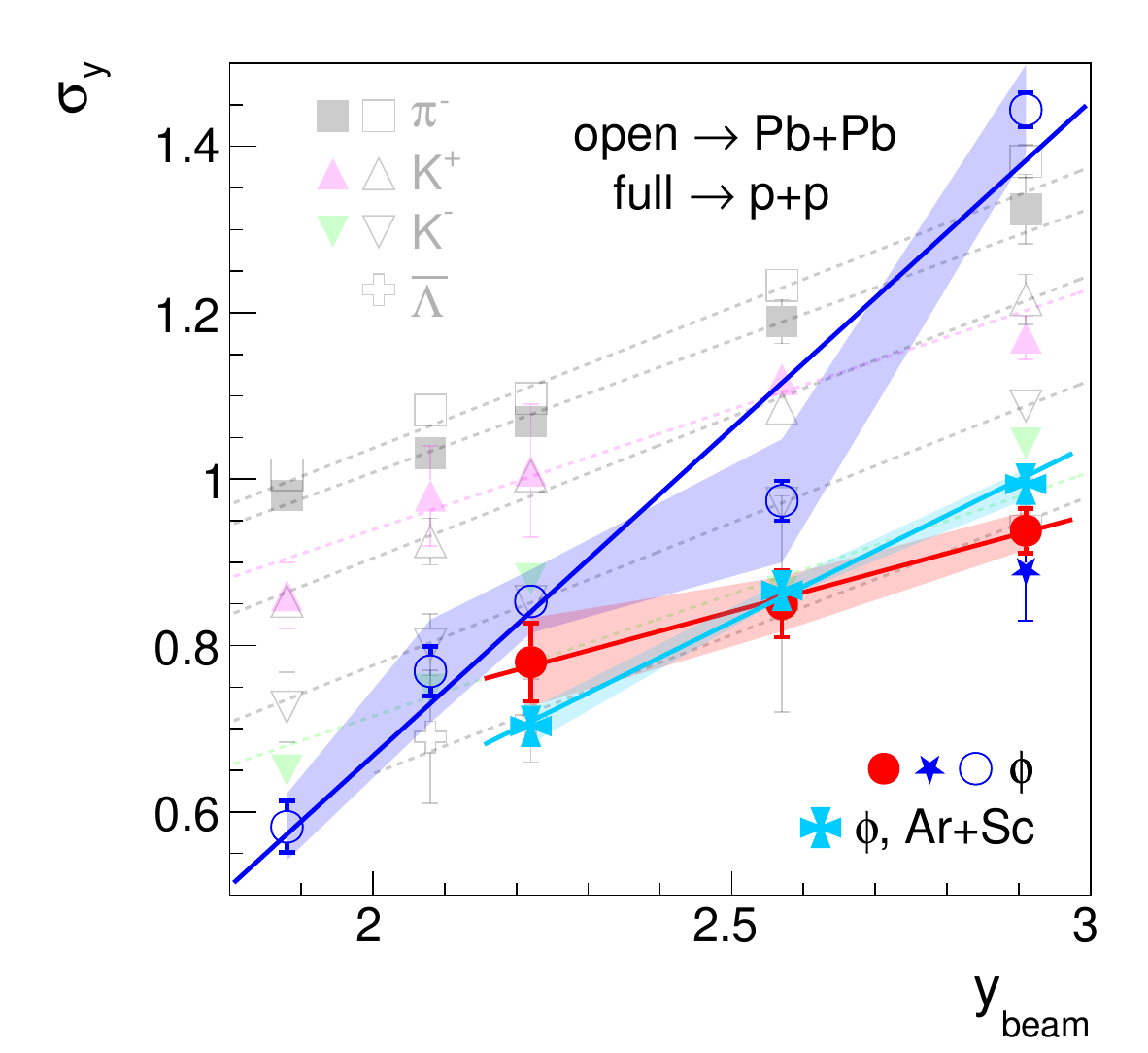}
  \caption{Widths of rapidity distributions for various particles as a function
  of beam rapidity. Data come from three different systems: central Pb+Pb~\cite{PhysRevC.66.054902, PhysRevC.77.024903, PhysRevLett.93.022302, PhysRevC.78.044907}, minimum bias \pp~\cite{Abgrall_2014, Pulawski:2240156, NA61SHINE:2019gqe, AFANASIEV200059} and central Ar+Sc. Vertical bars represent the statistical uncertainties, and the bands systematic ones.}
  \label{fig:rapidityWidht}
\end{SCfigure}

Fig.~\ref{fig:rapidityWidht} shows the widths of the rapidity distributions as a
function of the beam rapidity for various particles from three different
systems. Pale background consist of $\pi^-$, $K^+$, $K^-$ and
$\overline{\Lambda}$ produced in Pb+Pb~\cite{PhysRevC.66.054902,
PhysRevC.77.024903, PhysRevLett.93.022302} and \pp~\cite{Abgrall_2014,
Pulawski:2240156} collisions, which seem to follow a similar trend with
increasing beam rapidity. Overlaid on this background are the widths of the
$\phi$ rapidity distributions for Pb+Pb~\cite{PhysRevC.78.044907}, \pp~\cite{NA61SHINE:2019gqe, AFANASIEV200059}, and Ar+Sc (this analysis). The points representing $\phi$ from Pb+Pb deviate from the overall trend, whereas the $\phi$ from \pp does not. Therefore, examining $\phi$ production in intermediate systems, such as Ar+Sc, may help to clarify this puzzling behaviour. As visible in Fig.~\ref{fig:rapidityWidht}, the widths for $\phi$ mesons from Ar+Sc collisions are very similar to those from \pp reactions.
\sidecaptionvpos{figure}{c}

\begin{figure}[] \centering \includegraphics[page=1, width=0.329\textwidth]{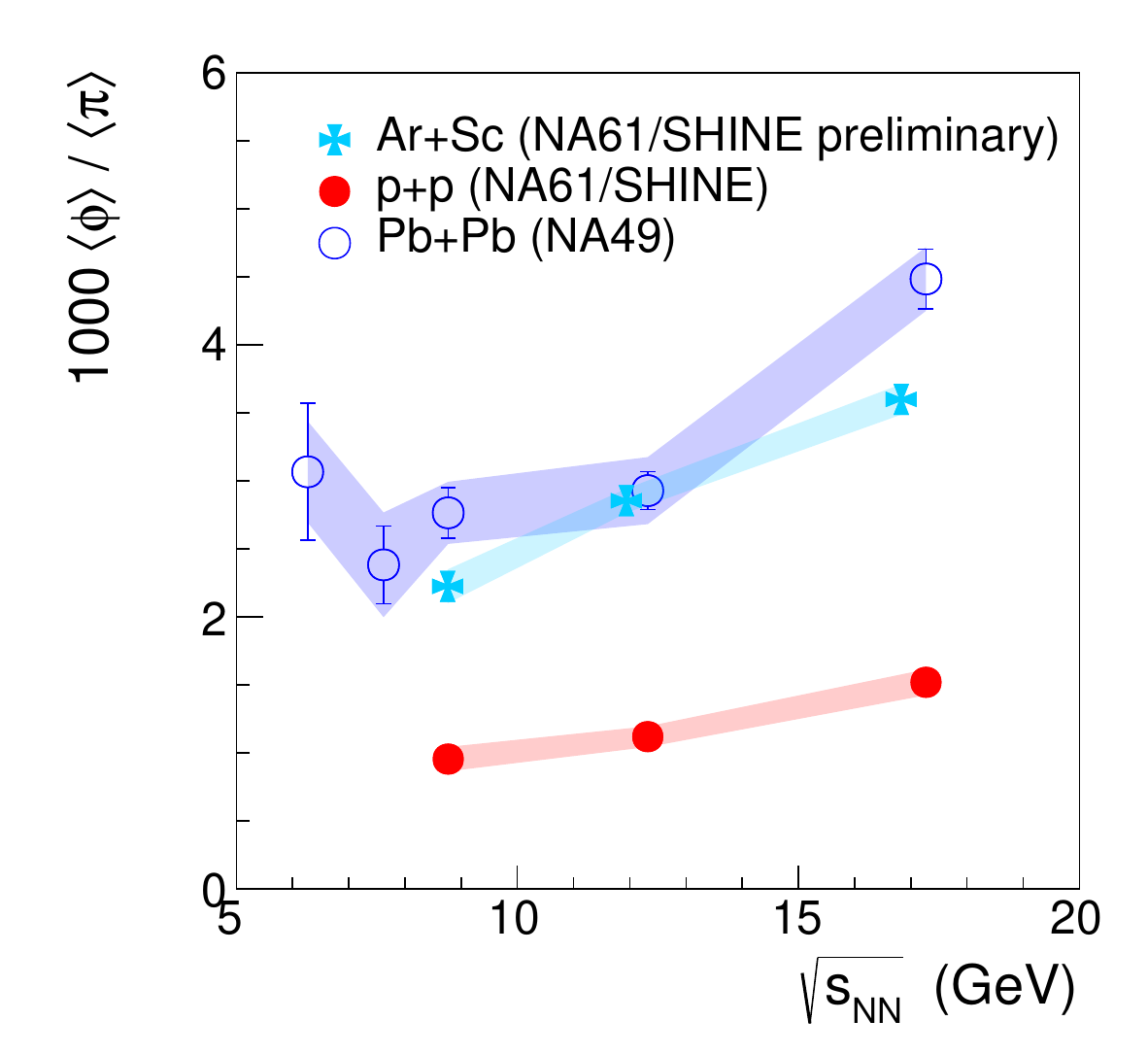} \includegraphics[page=2, width=0.329\textwidth]{figures/yield_ratios.pdf} \includegraphics[page=3, width=0.329\textwidth]{figures/yield_ratios.pdf} \caption{Left: energy dependence of the $\phi / \pi$ ratio for the minimum bias \pp, central Ar+Sc, and central Pb+Pb collisions. Middle and right: enhancement of $\phi$ and charged kaon production relative to pion production in Pb+Pb~\cite{PhysRevC.66.054902, PhysRevC.78.044907} and Ar+Sc~\cite{NA61SHINE:2023epu} collisions over the \pp~\cite{NA61SHINE:2017fne, NA61SHINE:2019gqe} reactions, as a function of the collision energy. Vertical bars denote statistical uncertainties, while shaded bands represent systematic uncertainties.} \label{fig:yieldRatios} \end{figure}

The left panel of Fig.~\ref{fig:yieldRatios} presents the energy dependence of the $\phi / \pi$ ratio for three systems: \pp, Ar+Sc, and Pb+Pb. The ratio for Ar+Sc is slightly lower than for Pb+Pb but much higher than for \pp collisions. The middle panel shows the double ratios for $\phi/\pi$, defined as:
\begin{equation}
\text{double ratio }(\text{Pb+Pb}) / (p+p) =
  \frac
  {(\langle \phi \rangle/\langle \pi \rangle)_{\text{Pb+Pb}}}
  {(\langle \phi \rangle/\langle \pi \rangle)_{p+p}}\,.
\end{equation}
This quantity is defined correspondingly for other particles. The right panel of Fig.~\ref{fig:yieldRatios} also displays the double ratio, but instead of comparing Pb+Pb, it compares Ar+Sc to \pp.  The enhancement in both Ar+Sc and Pb+Pb reactions over the \pp collisions, of particle production relative to pion production, is slightly higher for $\phi$ mesons than that for kaons, and is independent of the collision energy within the considered range. The data for \pp collisions is taken from Refs.~\cite{NA61SHINE:2017fne, NA61SHINE:2019gqe}, for Pb+Pb from Refs.~\cite{PhysRevC.66.054902, PhysRevC.78.044907} and for Ar+Sc from Refs.~\cite{NA61SHINE:2023epu}.  

\section{Summary}
\label{summary}
Preliminary results on the $\phi$(1020) meson production in central Ar+Sc reactions at beam momenta of 150$A$, 75$A$ and 40$A$ GeV/$c$ were shown.  They include double differential $(y, p_T)$ spectra, $p_T$-integrated rapidity distributions, and total yields. The rapidity widths of $\phi$ mesons produced in Ar+Sc collisions are similar to those from \pp collisions and different than in Pb+Pb. On the other hand, the ratio of $\phi$ to $\pi$ mesons production in Ar+Sc collisions is slightly lower than for Pb+Pb reactions and significantly higher than for \pp. Additionally, the enhancement of $\phi/\pi$ mesons production in both Ar+Sc and Pb+Pb collisions over the \pp, is slightly higher than that for charged kaons relative to pions and independent of the collision energy in the considered range.

This work was supported by the Polish National Agency for Academic Exchange NAWA under the Programme STER - Internationalisation of doctoral schools, Project no.  PPI/STE/2020/1/00020

\bibliography{bibliography.bib}

\end{document}